\documentclass[apjl]{emulateapj}
\usepackage{apjfonts}

\slugcomment{Astrophysical Journal Letters, in press (Vol.~650, 2006 October)}

\shorttitle{$M$--$\sigma$ FOR NUCLEAR CLUSTERS}
\shortauthors{McLAUGHLIN ET AL.}

\begin{document}

\title{The $M$--$\sigma$ Relation for Nucleated Galaxies}

\author{Dean E. McLaughlin, Andrew R. King, and Sergei Nayakshin}
\affil{Department of Physics and Astronomy, University of
Leicester, Leicester LE1 7RH, UK \\
{\sl dean.mclaughlin@astro.le.ac.uk ; ark@astro.le.ac.uk ;
sergei.nayakshin@astro.le.ac.uk}}

\begin{abstract}
Momentum feedback from super-Eddington accretion offers a simple
explanation for the observed $M$--$\sigma$ and $M$--$M_{\rm spher}$
relations between supermassive black holes (SMBH) and the spheroids of
their host galaxies. Recently \citeauthor{FerrareseEtal06} and
\citeauthor{WehnerHarris06} observed analogous relations
between the masses of central star clusters and their hosts. We show
that stellar winds and supernovae from such nuclear clusters (NC)
give similar feedback explanations for this case also, and we discuss
the connection to the Faber-Jackson relation for the spheroids themselves.
\end{abstract}

\keywords{galaxies: star clusters, formation, nuclei} 

\section{Introduction}
\label{sec:intro}

It is well known that the masses of
supermassive black holes (SMBHs) in the nuclei of early-type galaxies
and late-type bulges correlate tightly with the velocity dispersions of the
stellar spheroids: $M_{\rm BH} \propto \sigma^{x}$, with
$x \simeq 4.0$--4.5 \citep{TremaineEtal02,FerrareseFord05}.
$M_{\rm BH}$ also increases nearly linearly with galaxy spheroid mass
(e.g., \citealt{HaringRix04}). It is now becoming clear that many galaxies
have nuclear star clusters (NCs) with masses similarly connected to the host
properties.

A recent HST/ACS survey of 100 early-type galaxies in Virgo
has found that $\sim\! 70\%$--80\% of systems with
$-20.5\la M_B\la -15$ contain dense nuclear components that are
resolved by HST into star clusters with luminosities
$L\simeq 10^5\,L_\odot$--$5\times10^7\,L_\odot$,
half-light radii of about 4~pc
(possibly with a size-luminosity relation among
the brighter nuclei), and
colors suggesting ages of $\sim\! 2$--10~Gyr and metallicities
${\rm [Fe/H]}\approx -0.5 \pm 1$ \citep{CoteEtal06}.
Most spiral bulges and bulgeless disk galaxies also
contain nuclear clusters with rather similar properties
\citep{PhillipsEtal96, CarolloEtal98, BoekerEtal02,
WalcherEtal05}.

All known NCs have masses less than a few $10^8\,M_\odot$, while
most measured SMBHs have
$M_{\rm BH}\ga 10^8\,M_\odot$. The apparent paucity of low-mass
SMBHs is at least partly a selection effect, but the upper limit on the
nuclear clusters may well be real.
\citet{CoteEtal06} find that NC luminosity increases with spheroid
luminosity, such that $M_{\rm NC}\ga 2\times10^8\,M_\odot$ is expected for
galaxies with $M_B\la -20.5$. But although they have HST surface
photometry of the cores of {\it all} such galaxies in Virgo,
\citeauthor{CoteEtal06} find no evidence for
nucleation in any of them.

\citet{FerrareseEtal06} have obtained long-slit
spectra for 29 of the nucleated ellipticals in Virgo.
They find that the NC masses correlate well with
the galaxies' velocity dispersions
averaged over an effective radius ($\approx1$--2 kpc):
$M_{\rm NC}\propto \sigma^x$ with $x=4.3\pm0.6$, essentially the same
as for the SMBH relation.
However, they also find an {\it offset} between
the cluster and black-hole scalings.
Fitting power-law $M$--$\sigma$ relations with
$x\equiv 4$ to the \citeauthor{FerrareseEtal06} NC data and
to SMBH data from the literature yields (L.~Ferrarese, private communication)
\medskip
\begin{equation}
\begin{array}{lcl}
\log\,M_{{\rm NC}, 8} & = & (1.25 \pm 0.55) \ + \ 4\,\log\,\sigma_{200}  \\
\log\,M_{{\rm BH}, 8} & = & (0.25 \pm 0.33) \ + \ 4\,\log\,\sigma_{200}  \ ,
\end{array}
\label{eq:msigma4}
\end{equation}

\medskip
\noindent
where $M_8\equiv M/10^8\,M_\odot$
and $\sigma_{200}\equiv \sigma/(200\ {\rm km\,s}^{-1})$.
A limit of $M_{\rm NC}\la 2\times10^8\,M_\odot$ thus
corresponds to $\sigma\la 120$ km~s$^{-1}$. The nuclei of spheroids
with velocity dispersion less than this are dominated by stellar
clusters, with the mass of any SMBH that might be present expected to be
$\simeq\! 10$ times smaller. More massive galaxies apparently
always contain nuclear SMBHs but, so far as is known, not NCs.

\citet{WehnerHarris06} use photometric data in the literature for
about 40 dwarf-elliptical nuclei to show that 
$M_{\rm NC}$ increases almost linearly with galaxy spheroid mass.
\citet{FerrareseEtal06} also find this for
their nucleated Virgo galaxies with dynamical mass estimates.
Moreover, both studies
conclude that the $M_{\rm NC}$--$M_{\rm spher}$ and
$M_{\rm BH}$--$M_{\rm spher}$ relations meet almost seamlessly at a
mass scale $\sim\! 10^8\,M_\odot$, i.e., there is no large offset as in the
$M$--$\sigma$ relations.
These authors therefore refer to nuclear clusters and supermassive black
holes together as ``central massive objects,'' or
CMOs. We adopt this term here.

A derivation of the $M_{\rm BH}$--$\sigma$ scaling
has been given by \citet{King03, King05} (see
also \citealt{Fabian99}; \citealt{MurrayEtal05};
\citealt{BegelmanNath05}). He considers super-Eddington
accretion on to a seed SMBH at the center of an isothermal
dark-matter halo.
Accretion feedback produces a momentum-driven superbubble
that sweeps ambient gas into a thin shell, which expands into the galaxy.
Eventually the shock cooling time becomes so long that the shell
becomes energy-driven and accelerates to escape the
galaxy. This truncates accretion and
freezes in a relation of the form $M_{\rm BH}\propto \sigma^4$ that,
with no free parameters, matches the observed one remarkably well.
A roughly linear relation between $M_{\rm BH}$ and $M_{\rm spher}$ is also
established as part of this process.

In this {\it Letter}, we examine the possibility that the mass of a central
star cluster in a protogalaxy might be similarly self-regulated, by
feedback from stellar winds and supernovae.\footnotemark
\footnotetext{After this paper was submitted, a preprint appeared by
  \citet{LiEtal06}, which uses numerical simulations to study the development
  of an apparently common $M_{\rm CMO}$--$M_{\rm spher}$ relation for both NCs
  and SMBHs. Unlike that work, in our analytical model we consider
  stellar feedback explicitly to explain this result, and moreover we
  attempt to understand the {\it offset} between the $M$--$\sigma$
  scalings for NCs vs.~SMBHs.}

\section{The $M_{\rm CMO}-\sigma$ Relation}
\label{sec:msigma}

The argument of \citet{King03,King05} for the $M_{\rm BH}$--$\sigma$ relation
has gas in a protogalaxy flowing in to a low--mass, seed black hole at
super--Eddington rate. This grows the mass of the hole, but also drives an
intense outflow with momentum flux given by the Eddington luminosity:
$\dot{M} v_w \simeq L_{\rm Edd}/c$, independent of the actual
supercritical accretion rate \citep{KingPounds03}. Here $v_w$ is the outflow
velocity and $L_{\rm Edd}=4\pi G M_{\rm BH} c/\kappa$, with $\kappa=0.398$
cm$^2$~g$^{-1}$ the electron scattering opacity.

\citet{King03} shows that this outflow is initially momentum-conserving,
as the shocked gas cools efficiently. The ambient medium is swept
up into a thin supershell, which is driven outwards by the ram
pressure of the SMBH wind:
$\rho_w v_w^2=\dot{M} v_w/(4\pi R^2) =G M_{\rm BH}/(\kappa R^2)$
at radius $R$.
The dark-matter halo is assumed to be an isothermal sphere and
the ambient gas fraction spatially constant,
so that $\rho_{\rm amb}(R) = f_g \sigma^2 / (2 \pi G R^2)$.
Realistically, there could be a gradient in $f_g$ due to a concentration
of cool gas towards the center of the galaxy, but here we take $f_g$ to
be everywhere equal to its average over the entire halo:
$f_g = \Omega_b/\Omega_m = 0.16$ \citep{SpergelEtal03}.
If gravity is ignored, the supershell accelerates once $M_{\rm BH}$ has
grown to the point that $\rho_w v_w^2 \ga \rho_{\rm amb}\sigma^2$.
Any ambient gas outside it is then driven out of the galaxy, stopping
the growth of the SMBH. This happens when $M_{\rm BH}=f_g \kappa
\sigma^4/(2\pi G^2)$.

Including the gravity of dark matter inside the
superbubble alters this result by a factor of two \citep{King05}.
The initial dynamical expansion of the shell then stalls at a
radius $R_{\rm stall} \sim (1-M_{\rm BH}/M_{\rm crit})^{-1}$, with
\medskip
\begin{equation}
M_{\rm crit} = f_g \kappa \sigma^4/(\pi G^2)\ ,
\label{eq:Mcmo1}
\end{equation}

\medskip
\noindent
where the gravity balances ram pressure. So long as
$M_{\rm BH} \ll M_{\rm crit}$, this happens well inside the galaxy.
More gas can then filter through to the nucleus, feeding the hole and causing
the shell
to re-expand (on a Salpeter timescale) to a larger
$R_{\rm stall}$ appropriate to the new $M_{\rm BH}$. As $M_{\rm BH}$
approaches $M_{\rm crit}$,
however, the stall radius becomes very large, and before the
shell can actually reach it the gas cooling time becomes longer than
the crossing time of the bubble. The shell then enters an energy-conserving
snowplow phase and accelerates to escape the galaxy,
leaving $M_{\rm BH}\simeq M_{\rm crit}$. As emphasized above, it is
noteworthy that $M_{\rm crit}$ contains no free parameter.

Our main point is that the above argument is
qualitatively unchanged if the CMO is not an SMBH but instead a {\it
very young} star cluster where massive stars are still present.
Then stellar winds and supernovae drive a superwind from the
nucleus with a momentum flux that is much less than $L_{\rm Edd}/c$
but still directly proportional to it. We can thus treat the
two types of CMO simultaneously by parametrizing the wind thrust as
\medskip
\begin{equation}
\dot{M} v_w \equiv \lambda \ L_{\rm Edd}/c
              =    \lambda \ (4\pi G M_{\rm CMO}/\kappa) \ .
\label{eq:pflux}
\end{equation}

\medskip
\noindent
Here $\lambda$ takes a value $\simeq\! 1$ in the black hole case,
but a value $\ll\! 1$ (related to the mass fraction in massive stars) for a
nuclear cluster.
The limiting mass in equation (2) becomes
\medskip
\begin{equation}
M_{\rm CMO} = 3.67\times 10^8 \ M_\odot\ 
               \lambda^{-1}\, \sigma_{200}^4\, (f_g/0.16) \ ,
\label{eq:Mcmo2}
\end{equation}

\medskip
\noindent
and the offset $M_{\rm BH}$--$\sigma$ and
$M_{\rm NC}$--$\sigma$ relations of equation (\ref{eq:msigma4})
follow immediately if $\lambda\sim 0.1$ for a typical NC.

\section{Efficiency of Stellar Feedback}
\label{sec:lambda}

To evaluate the efficiency $\lambda$ of the massive--star feedback
from a young nuclear cluster, we rewrite equation (\ref{eq:pflux}) as
\medskip
\begin{equation}
\lambda_{\rm NC} = \frac{\dot{M} v_w}{4\pi G M_{\rm NC}/\kappa}
  = \frac{\dot{M} v_w}{4.2\times10^{27}\,(M_{\rm NC}/M_\odot)\ {\rm dyn}}
\label{eq:lambda}
\end{equation}

\medskip
\noindent
and estimate the separate contributions
from supernovae and stellar winds.

First, the combined momentum flux from all supernovae is
$2 N_{\rm SN} E_{\rm SN}/(v_{\rm SN} \tau_{\rm SN})$,
where $N_{\rm SN}\approx 0.011\,(M_{\rm NC}/M_\odot)$
is the number of stars with
mass $>\!\!8\,M_\odot$ in a cluster with a 
\citet{Chabrier03} IMF; $E_{\rm SN}=10^{51}$~erg is the energy released
per  supernova; $v_{\rm SN}\approx 4000$~km~s$^{-1}$ is the typical
ejecta velocity \citep{WeilerSramek88}; and
$\tau_{\rm SN}\approx2\times10^7$~yr is the main-sequence lifetime of an
``average'' SN progenitor \citep[see][]{LeithererEtal92}. Putting this
into equation (\ref{eq:lambda}) gives $\lambda_{\rm SN} \approx 0.02$.

Second, the line-driven wind from a single hot star produces a momentum
flux of $\approx\!\! (L_*/c)$ on average---somewhat less than this if
there are few lines to drive the wind, but several times higher for O and
Wolf-Rayet stars in which photons are multiply scattered
\citep[see, e.g.,][chapter 8]{LamersCassinelli99}. Using the
main-sequence mass-luminosity relation of \citet{ToutEtal96} to integrate
$(L_*/c)$ over all stars more massive than $5\,M_\odot$ in the IMF of
\citet{Chabrier03}, we find that the total momentum flux from stellar
winds is about $1.3\times 10^{26}\,(M_{\rm NC}/M_\odot)$~dyn, and
thus $\lambda_{\rm winds} \approx 0.03$.

Despite its very simple derivation, our final
\medskip
\begin{equation}
\lambda_{\rm NC} = \lambda_{\rm SN} + \lambda_{\rm winds} \approx 0.05
\label{eq:lambdavalue}
\end{equation}

\medskip
\noindent
is in good agreement with the values implied by the detailed
calculations of \citet{LeithererEtal92} for the total momentum deposition in
solar-metallicity starbursts. Note also that the stellar luminosity
corresponding to the limiting NC mass in equation (\ref{eq:Mcmo2}) with
$\lambda=0.05$ is comparable to that derived by \citet{MurrayEtal05}
from related considerations (see their eq.~[18]).

One caveat here is that, while SN momentum fluxes are insensitive to stellar
metallicity, wind momenta are roughly proportional to $Z$
\citep{LeithererEtal92}. For $(Z/Z_\odot)\approx 1/3$,
typical of NCs in Virgo, this might then
imply a net $\lambda_{\rm NC}\sim 0.03$.
However, this effect could be easily balanced by increases in both
$\lambda_{\rm SN}$ and $\lambda_{\rm winds}$ if the IMF in these dense,
central starbursts were slightly ``top-heavy,'' as may be the case near the
center of the Milky Way \citep[e.g.,][]{NayakshinSunyaev05,StolteEtal05}. We
proceed assuming the fiducial value for $\lambda_{\rm NC}$
in equation (\ref{eq:lambdavalue}).

With $\lambda\approx0.05$ for a nuclear cluster and $\lambda\simeq 1$
for a supermassive black hole, equation (\ref{eq:Mcmo2}) implies an offset of
a factor of 20 between the two $M_{\rm CMO}$--$\sigma$
relations, while observationally it is only a factor of 10
(eq.~[\ref{eq:msigma4}]). However, population-synthesis models
\citep[e.g.,][]{FiocRoccaVolmerange97,BruzualCharlot03}
show that a star cluster more than $\sim\! 10^9$~yr old will have lost some
40\%--50\% of its initial total mass to stellar winds
and supernovae, and to the conversion of massive
stars into degenerate remnants. Thus, we expect the
$M_{\rm NC}$--$\sigma$ scaling originally to have been
more offset from the SMBH correlation than it is now,
by an additional factor of about 2.
By the same reasoning, the fact that the two
$M_{\rm CMO}$--$M_{\rm spher}$ relations currently appear to have
very similar normalizations must be something of a coincidence.

\section{The $M_{\rm CMO}-M_{\rm spher}$ Relation}
\label{sec:mmbulge}

In this feedback--regulated picture of CMO and galaxy formation, an
$M_{\rm CMO}$--$\sigma$ relation emerges as the primary correlation.
A relation between
$M_{\rm CMO}$ and spheroid mass follows
by combining equation (\ref{eq:Mcmo2}) with details of the cooling
of the wind from the central object. The basic steps are outlined in
\citet{King03}. Knowing how the shocked gas cools,
we find the cooling timescale $t_{\rm cool}$ as a
function of supershell radius $R$ and compare it to the dynamical time
$t_{\rm flow}=R/v_{\rm shell}$. For small radii, $t_{\rm cool}<t_{\rm flow}$
and the outflow is momentum--driven.
However, when the CMO is at about the critical mass in equation
(\ref{eq:Mcmo2}), the bubble is so large that $t_{\rm cool}$
exceeds $t_{\rm flow}$, the thin shell becomes energy-conserving,
and the wind can escape the galaxy. We use $R_{\rm cool}$ to denote the
radius at which this happens. The detailed fate
of the swept-up ambient gas afterwards is beyond the scope of this
{\it Letter}, but in general terms it should re-collapse to much
smaller radii (since the CMO wind that pushed the gas to large $R$ in the
first place carries essentially no angular momentum). It will cool
rapidly as it does, because $t_{\rm cool}<t_{\rm flow}$ by construction
inside $R<R_{\rm cool}$, and $t_{\rm flow}$ is of order the
free-fall time in the halo. We therefore expect most of the ambient gas
in the supershell at the point of wind blow-out to form a concentrated
stellar spheroid, and thus we identify the mass of the shell at $R_{\rm cool}$
with $M_{\rm spher}$.

\citet{King03} shows that for a relativistic wind from an SMBH, the
swept-up gas cools by Compton scattering, and ultimately,
\medskip
\begin{equation}
\frac{M_{\rm BH}}{M_{\rm spher}}  = 1.6\times10^{-3} \ 
                  b^{4/5} (c/v_w)^{8/5}
                  (f_g/0.16)^{-3/5}
		  M_{{\rm spher}, 11}^{-1/5} \ .
\label{eq:MMbcomp}
\end{equation}

\medskip
\noindent
Here $v_w\sim c$; $b\sim 1$ is an outflow collimation parameter, and
$M_{\rm spher}$ is in units of $10^{11}\,M_\odot$. This agrees well
with the observed SMBH-to-spheroid mass ratios in giant galaxies
\citep{HaringRix04}.

If the CMO is a star cluster, equation (\ref{eq:MMbcomp}) no longer
applies because the wind driving the superbubble is far from
relativistic, and the Compton cooling time for the shocked gas exceeds
a Hubble time. The cooling in this case is by atomic transitions.
However, the $M_{\rm NC}$--$M_{\rm spher}$ relation still
has the basic form of equation (\ref{eq:MMbcomp}), i.e.,
$M_{\rm CMO}\propto M_{\rm spher}^{{\rm 4/5}}$ in all cases.

To see this, note first that we always have
$t_{\rm flow}=R/v_{\rm shell} = R/\sqrt{2} \sigma$ 
for the shell as it escapes.
In the SMBH case,
the Compton cooling time is proportional to the inverse of the
radiation energy density, which is diluted by the $1/R^2$ law: $t_{\rm
cool}\propto (L_{\rm Edd}/4\pi R^2 c)^{-1} \propto R^2/M_{\rm BH}$.  In
the NC case, the cooling time has the
same dependence because the wind density {\it also} falls off as
$1/R^2$: $\rho_w \propto M_{\rm NC}/(4\pi R^2 v_w^2)$, and then
$t_{\rm cool}\propto \rho_w^{-1} \propto R^2/M_{\rm NC}$.
Thus, either type of CMO has $t_{\rm cool}=t_{\rm flow}$ at a radius
$R_{\rm cool} \propto M_{\rm CMO}/\sigma$, or simply
$R_{\rm cool} \propto \sigma^3$ using equation (\ref{eq:Mcmo2}). Finally,
$M_{\rm spher} \propto \sigma^2 R_{\rm cool} \propto \sigma^5$
and hence
$M_{\rm CMO} \propto M_{\rm spher}^{4/5}$.

In detail, the flow time is always \citep[e.g.,][]{King03}
\medskip
\begin{equation}
t_{\rm flow} = 6.6\times 10^6 \ {\rm yr}\ \ 
                   R_{\rm kpc} \, \sigma_{200} 
		   \, \lambda^{-1/2} M_8^{-1/2} (f_g/0.16)^{1/2}
\label{eq:tflow}
\end{equation}

\medskip
\noindent
for the shell radius $R$ in units of kpc and
$M_8\equiv M_{\rm CMO}/10^8\,M_\odot$. Again, $\lambda=1$ for an SMBH and
$\lambda \approx 0.05$ for an NC. In the latter case, we further find for
the radiative cooling time,
\medskip
\begin{equation}
t_{\rm cool} \simeq 1.4 \times10^4 \ {\rm yr}\ \ 
         R_{\rm kpc}^2 \, v_{w, 300}^{5.5}
         \, \lambda^{-1} \, M_8^{-1} \, (Z/Z_\odot)^{-0.6} \ ,
\label{eq:tcool}
\end{equation}

\medskip
\noindent
where $v_{w,300}$ is the speed of the cluster superwind in units of
300~km~s$^{-1}$. This follows from the definition
$t_{\rm cool}=\mu m_H k T/(\rho_w \Lambda_N)$, with $\mu\simeq0.6$;
the wind density $\rho_w$ given by the continuity equation,
$4\pi R^2\rho v_w^2 = \dot{M} v_w$;
the shock temperature $kT/\mu m_H = (3/16) v_w^2$; and $\Lambda_N$ the
normalized cooling function calculated by \citet{SutherlandDopita93}. This
last is approximately $\Lambda_N \simeq 3.55\times
10^{-18}\,(Z/Z_\odot)^{0.6}\,T^{-0.75}$ erg~cm$^3$~s$^{-1}$
for  $(Z/Z_\odot) = 0.1$--1 and $T\simeq 0.5$--$5\times 10^6$~K
($v_w\simeq200$--600 km s$^{-1}$).

Finding the radius at which $t_{\rm cool}=t_{\rm flow}$ leads to
$M_{\rm spher} = 2 f_g \sigma^2 R_{\rm cool}/G$, and combining with
equation (\ref{eq:Mcmo2}) gives
\medskip
\begin{equation}
\frac{M_{\rm NC}}{M_{\rm spher}} = 2.7\times10^{-4} \, \lambda^{-1} 
    \left(\frac{Z}{Z_\odot}\right)^{-0.48}
    \left(\frac{f_g}{0.16}\right)^{-3/5}
    v_{w,300}^{4.4} \, M_{{\rm spher}, 11}^{-1/5} \ .
\label{eq:MMbatom}
\end{equation}

\medskip
\noindent
Comparing equation (\ref{eq:MMbcomp}) to equation
(\ref{eq:MMbatom}) with $\lambda=0.05$, the predicted offset
of the {\it original} $M_{\rm NC}$--$M_{\rm spher}$ relation from the
corresponding SMBH relation is only a factor of $\approx\! 3$--4 for a
wind velocity near 300 km~s$^{-1}$ and slightly subsolar
metallicities. Allowing for the long-term mass loss from the NC,
discussed at the end of \S\ref{sec:lambda}, the normalizations of
the present-day scalings should agree, rather
fortuitively, to within a factor of two---just as
\citet{FerrareseEtal06} and \citet{WehnerHarris06} infer
observationally.

The contrast between this and the much larger offset separating
the two $M_{\rm CMO}$--$\sigma$ relations 
is due to the fact that, for given $\sigma$, the
radiative $t_{\rm cool}$ with an NC at the limiting mass of equation
(\ref{eq:Mcmo2}) is nearly 10 times shorter than the Compton cooling
time that applies when the CMO is an SMBH
\citep[see][]{King03}. Thus, the final $R_{\rm cool}$ and $M_{\rm
spher}$ are larger by this amount for a galaxy containing a nuclear star
cluster vs.~one with the same $\sigma$ but a central black hole.

\section{The Faber--Jackson Relation}
\label{sec:bulges}

For a galaxy with a supermassive black hole in its nucleus, equations
(\ref{eq:MMbcomp}) and (\ref{eq:Mcmo2}) imply a relation between the
mass and velocity dispersion of the spheroid alone \citep{King05}:
\medskip
\begin{equation}
M_{\rm spher}({\rm SMBH}) =
             2.8\times10^{11} \ M_\odot\ b^{-1} (c/v_w)^{-2}
                      (f_g/0.16)^2 \, \sigma_{\rm 200}^5 \ .
\label{eq:Mbcomp}
\end{equation}

\medskip
\noindent
Perhaps despite appearances, this prediction is consistent
with the well-known relation between the velocity dispersion and {\it
luminosity} of giant ellipticals, $L\propto \sigma^4$
\citep{FaberJackson76}. This is because the mass-to-light ratio
in the cores of these galaxies increases systematically
as the $\simeq\! 0.2$--$0.3$ power of luminosity
\citep{vanderMarel91, CappellariEtal06}, so that
the Faber-Jackson relation in fact implies
$M_{\rm spher}\propto \sigma^{5}$.
The normalization in equation (\ref{eq:Mbcomp}) is also in remarkably good
agreement with observation: from \citet{HaseganEtal05},
\medskip
\begin{equation}
M_{\rm spher}(\rm obs) \simeq 1.93\times10^{11}\,M_\odot
                                \ \sigma_{200}^{5.2}
\label{eq:Mbobs}
\end{equation}

\medskip
\noindent
for galaxies with $\sigma\ga 100$~km~s$^{-1}$---which, as we
discussed in \S\ref{sec:intro}, are the ones that contain SMBHs rather
than NCs.

For smaller galaxies with nuclei
dominated by star clusters, $t_{\rm cool}=t_{\rm flow}$ gives
the cooling radius
\medskip
\begin{equation}
R_{\rm cool}({\rm NC}) = 880 \ {\rm kpc}\ (Z/Z_\odot)^{0.6} (f_g/0.16)
                   \, v_{w,300}^{-5.5} \, \sigma_{200}^3 \ .
\label{eq:Ratom}
\end{equation}

\medskip
\noindent
As we mentioned above, this is nearly 10 times larger than the equivalent
scale in the SMBH-dominated case, and the spheroid
mass is consequently larger than in equation (\ref{eq:Mbcomp}):
\medskip
\begin{equation}
M_{\rm spher}({\rm NC}) =
    2.6\times10^{12} \ M_\odot\ (Z/Z_\odot)^{0.6} (f_g/0.16)^2
                      \, v_{w,300}^{-5.5} \, \sigma_{200}^5 \ .
\label{eq:Mbline}
\end{equation}

\medskip

In a plot of $\sigma$ vs.~$M_{\rm spher}$, we would therefore expect
low-mass, nucleated galaxies to define a locus with
$\sigma\propto M_{\rm spher}^{0.2}$, parallel to the 
non-nucleated galaxies with SMBHs but falling below them by a factor of
$\approx\!10^{-0.2}$.
Then, although our model does not predict a value for the final
effective radius of the stellar spheroid in terms of $R_{\rm cool}$,
the virial theorem requires any such scale to depend
on the spheroid mass roughly as $R_{\rm eff} \propto M_{\rm spher}^{0.6}$,
with nucleated galaxies lying above non-nucleated ones by a factor of
about 2.5.

\section{Discussion}
\label{sec:discussion}

Current data probably do not rule out the idea that nuclei of galaxies with
$\sigma\la 120$ km~s$^{-1}$ could harbor {\it both} star clusters {\it
and} SMBHs some $\sim\!10$ times less massive, and thus it will be
important to ask how such galaxies might choose between SMBH and NC
feedback channels in regulating their formation.
For now, a possibly more straightforward question is why there are {\it no}
nuclear clusters in larger galaxies with
$\sigma\ga120$ km~s$^{-1}$ ($M_{\rm spher}\ga 2$--$3\times10^{10}\,M_\odot$),
which apparently all contain black holes.

For our basic scenario to be
self-consistent, we evidently require that the dynamical time of the
superbubble always be shorter than that of the entire
halo, until the point of blow-out when
$t_{\rm flow}=t_{\rm cool}$. Equivalently, the radius of the shell must
always be less than the halo virial radius, and in particular
$R_{\rm cool}\la R_{\rm vir}$ is required for the growth of a CMO (whether an
SMBH or an NC) to be self-regulated as we envision. From the relations in
\citet{BryanNorman98}, if $\Omega_{m,0}=0.3$,
$\Omega_\Lambda=0.7$, and $H_0=70$~km~s$^{-1}$~Mpc$^{-1}$, then
$R_{\rm vir}\approx540\ {\rm kpc}\ \sigma_{\rm 200}\,(1+z)^{-1.1}$
(accurate to better than 10\% for $z\le 2$). Combining this with equation
(\ref{eq:Ratom}), our picture can work for NCs forming at redshift
$z_{\rm NC}$ only in halos with 
$\sigma \la 160\ {\rm km~s}^{-1}\,(1+z_{\rm NC})^{-0.55}$.

If nuclear clusters are typically  $\sim\! 5$ Gyr old,
$z_{\rm NC}\simeq 0.5$ and this upper limit
becomes $\sigma\la 130$~km~s$^{-1}$. In halos with
velocity dispersion higher than this, the superbubble blown by an
NC reaches the halo virial radius before it becomes
energy-conserving and can accelerate to escape. It is presumably then held
there, or even driven
into collapse, by the infall of material from beyond $R_{\rm
vir}$. The growth of the central star ``cluster'' is never choked
off, and it ultimately becomes indistinguishable from the galaxy spheroid
itself. When the CMO is a black hole, the much longer Compton-cooling
time found in \citet{King03} implies that $R_{\rm cool}\la R_{\rm vir}$ for
all $\sigma\la 500\,{\rm km~s}^{-1}\,(1+z_{\rm
BH})^{-0.55}$, so it is only in very massive systems indeed that
the self-regulated $M_{\rm BH}$--$\sigma$ relation of
equation (\ref{eq:Mcmo1}) breaks down.

\acknowledgements
We thank Laura Ferrarese and Pat C\^ot\'e for helpful discussions.
DEM is supported by a PPARC standard grant.
Theoretical Astrophysics at the University of Leicester is also
supported by a PPARC rolling grant. ARK gratefully acknowledges a
Royal Society Wolfson Research Merit Award.

\section*{}
\label{sec:note}

{\it Note added in proof.}---J.~Rossa et al.~(AJ, 132, 1074 [2006]) have
recently analyzed a sample of nuclear clusters in 40 spiral galaxies, finding
a relation between $M_{\rm NC}$ and the luminosity of the galaxies' {\it
  bulge} components, which has the same slope as the relation between SMBH
mass and bulge luminosity with a comparable intercept. The situation for NCs
in late-type spheroids thus appears very similar to that found by
\citet{WehnerHarris06} and \citet{FerrareseEtal06} to hold in early-type
galaxies.


\begin{thebibliography}{}

\bibitem[Begelman \& Nath(2005)]{BegelmanNath05} Begelman, M. C., \& Nath,
  B. B. 2005, \mnras, 361, 1387

\bibitem[B\"oker et al.(2002)]{BoekerEtal02} B\"oker, T., Laine, S., van der
  Marel, R. P., Sarzi, M., Rix, H.-W., Ho, L. C., \& Shields, J. C. 2002, \aj,
  123, 1389

\bibitem[Bruzual \& Charlot(2003)]{BruzualCharlot03} Bruzual, G., \& Charlot,
  S. 2003, \mnras, 344, 1000

\bibitem[Bryan \& Norman(1998)]{BryanNorman98} Bryan, G. L., \& Norman,
  M. L. 1998, \apj, 495, 80

\bibitem[Cappellari et al.(2006)]{CappellariEtal06} Cappellari, M., et
  al. 2006, \mnras, 366, 1126

\bibitem[Carollo, Stiavelli, \& Mack(1998)]{CarolloEtal98} Carollo, C. M.,
  Stiavelli, M., \& Mack, J. 1998, \aj, 116, 68

\bibitem[Chabrier(2003)]{Chabrier03} Chabrier, G. 2003, \pasp, 115, 763

\bibitem[C\^ot\'e et al.(2006)]{CoteEtal06} C\^ot\'e, P., et al. 2006, \apjs,
  165, 57

\bibitem[Faber \& Jackson(1976)]{FaberJackson76} Faber, S. M., \& Jackson,
  R. E. 1976, \apj, 204, 668

\bibitem[Fabian(1999)]{Fabian99} Fabian, A. C. 1999, \mnras, 308, L39

\bibitem[Ferrarese \& Ford(2005)]{FerrareseFord05} Ferrarese, L., \& Ford,
  H. C. 2005, Space Science Reviews, 116, 523

\bibitem[Ferrarese et al.(2006)]{FerrareseEtal06} Ferrarese, L., et
  al. 2006, \apj, 644, L21

\bibitem[Fioc \& Rocca-Volmerange(1997)]{FiocRoccaVolmerange97} Fioc, M., \&
  Rocca-Volmerange, B. 1997, \aap, 326, 950

\bibitem[H\"aring \& Rix(2004)]{HaringRix04} H\"aring, N., \& Rix, H.-W. 2004,
  \apj, 604, L89

\bibitem[Ha\c{s}egan et al.(2005)]{HaseganEtal05} Ha\c{s}egan, M., et
  al. 2005, \apj, 627, 203

\bibitem[King(2003)]{King03} King, A. R. 2003, \apj, 596, L27

\bibitem[King(2005)]{King05} King, A. R. 2005, \apj, 635, L121

\bibitem[King \& Pounds(2003)]{KingPounds03} King, A. R., \& Pounds,
  K. A. 2003, \mnras, 345, 657

\bibitem[Lamers \& Cassinelli(1999)]{LamersCassinelli99} Lamers,
  H. J. G. L. M., \& Cassinelli, J. P. 1999, Introduction to Stellar Winds
  (Cambridge: Cambridge University Press)  

\bibitem[Leitherer, Robert, \& Drissen(1992)]{LeithererEtal92} Leitherer, C.,
  Robert, C., \& Drissen, L. 1992, \apj, 401, 596

\bibitem[Li, Haiman, \& Mac Low(2006)]{LiEtal06} Li, Y., Haiman, Z., \& Mac
  Low, M.-M. 2006, {\tt astro-ph/0607444}

\bibitem[Murray, Quataert, \& Thompson(2005)]{MurrayEtal05} Murray, N.,
  Quataert, E., \& Thompson, T. A. 2005, \apj, 618, 569

\bibitem[Nayakshin \& Sunyaev(2005)]{NayakshinSunyaev05} Nayakshin, S., \&
  Sunyaev, R. 2005, \mnras, 364, L23 

\bibitem[Phillips et al.(1996)]{PhillipsEtal96} Phillips, A. C., Illingworth,
  G. D., MacKenty, J. W., \& Franx, M. 1996, \aj, 111, 1566

\bibitem[Spergel et al.(2003)]{SpergelEtal03} Spergel, D. N., et al. 2003,
  \apjs, 148, 175

\bibitem[Stolte et al.(2005)]{StolteEtal05} Stolte, A., Brandner, W., Grebel,
  E.~K., Lenzen, R., \& Lagrange, A.-M. 2005, \apj, 628, L113

\bibitem[Sutherland \& Dopita(1993)]{SutherlandDopita93} Sutherland, R. S., \&
  Dopita, M. A. 1993, \apjs, 88, 253

\bibitem[Tout et al.(1996)]{ToutEtal96} Tout, C. A., Pols, O. R., Eggleton,
  P. P., \& Han, Z. 1996, \mnras, 281, 257

\bibitem[Tremaine et al.(2002)]{TremaineEtal02} Tremaine, S., et al. 2002,
  \apj, 574, 740

\bibitem[van der Marel(1991)]{vanderMarel91} van der Marel, R. P. 1991,
  \mnras, 253, 710

\bibitem[Walcher et al.(2005)]{WalcherEtal05} Walcher, C. J., et al. 2005,
  \apj, 618, 237

\bibitem[Wehner \& Harris(2006)]{WehnerHarris06} Wehner, E. H., \& Harris,
  W. E. 2006, \apj, 644, L17

\bibitem[Weiler \& Sramek(1988)]{WeilerSramek88} Weiler, K. W., \& Sramek,
  R. A. 1988, \araa, 26, 295

\end{thebibliography}
\end{document}